\documentclass[twoside]{article}
\usepackage{fleqn,espcrc2,epsfig}
\usepackage{amsfonts}

\usepackage[figuresright]{rotating}

\def\NN{\mathbb N}
\def\ZZ{\mathbb Z}

\def\RR{\mathbb R}
\def\D{{\cal D}}
\def\g{ \mathfrak{g}}
\def\s{  \mathfrak{sl}}
\def\noi{\noindent}

\newcommand{\AmS}{{\protect\the\textfont2
  A\kern-.1667em\lower.5ex\hbox{M}\kern-.125emS}}
\title{Fractional Supersymmetry and Infinite Dimensional Lie Algebras}

\author{M. Rausch de Traubenberg
\address[MCSD]{ Universit\'e de Montpellier II,
Laboratoire de Physique Math\'ematique et Th\'eorique, \\
        Place E. Bataillon, 34095 Montpellier Cedex 05 , France}%
        \thanks{One lives from absence from Universit\'e Louis Pasteur,
        Laboratoire de Physique Th\'eorique, 67034 Strasbourg Cedex, France.}}
        
\begin{document}

\begin{abstract}
In an earlier work extensions of supersymmetry and super Lie algebras 
were constructed  consistently starting from any representation $\D$ 
of any Lie algebra $\g$. Here it is shown how infinite dimensional Lie 
algebras appear naturally within the framework of fractional supersymmetry. 
Using a differential realization of $\g$ this  infinite dimensional Lie 
algebra,  containing the Lie algebra $\g$ as a sub-algebra, is
explicitly constructed. 
\vspace{1pc}
\end{abstract}

\maketitle

\section{Introduction}

It is a pleasure  to write this contribution in memory of the 
75th anniversary of D. V. Volkov. Among other things he was a pioneer 
both in supersymmetry and alternative statistics \cite{volkov}. 
This paper can be seen as an attempt to combine  
those two concepts in a sense specified latter on.\\
Symmetry is always a powerful tool to understand properties of systems.
Space-time symmetry is central in particle physics.
When studying possible symmetries of space-time, two theorems put severe
constraints on allowed  space-time symmetries: the
Coleman and Mandula \cite{cm} and Haag, Lopuszanski and Sohnius \cite{hls} 
theorems. They lead to gauge theories and/or supersymmetry.
However, for those types of theories  bosonic or
fermionic representations of the Lorentz group are introduced.
 So, if one imagine
symmetries acting on representations which are neither bosonic nor
fermionic, one is, in principle able to define symmetries generalizing
supersymmetry. Several
possibilities have been considered in the literature 
\cite{ker}-\cite{fliealg}, 
the intuitive idea being
that the generators of the Poincar\'e algebra are obtained as an appropriate
product of more fundamental additional symmetries. These new generators
are in a representation of the Lorentz algebra which can be neither
bosonic nor fermionic (bosonic charges close under commutators and 
generate a Lie algebra, whilst fermionic charges close under anticommutators
and  induce  super-Lie algebras).  Here, we are  interested in one possible
extension of supersymmetry named fractional supersymmetry.

\section{$F-$Lie algebras}
The natural mathematical  structure, generalizing the concept of 
super-Lie algebras and relevant for the algebraic description of fractional
supersymmetry has been introduced in \cite{vg}  and was called 
fractional super-Lie algebra ($F-$Lie algebra for short). Here, we do not 
want  to go into the  detailed definition of this structure but only
recall the basic points, useful for our purpose. 
More details can be obtained in \cite{vg}. 
Basically, an $F-$Lie algebra is a $\ZZ_F-$graded vector space,

\begin{equation}
S= B \oplus A_1 \oplus \cdots \oplus A_{F-1},
\label{eq:F-lie}
\end{equation}

\noindent
which endows  the following structure:
\begin{enumerate}
\item $B$ is the graded zero part of $S$ and is a Lie algebra.
\item All the graded part of $S$, $A_i,i=1,\cdots, F-1$
(of grade $i$) are  appropriate
representations of $B$ (in general of infinite dimension {\it i.e.}
 a Verma module). 
\item Denoting ${ \cal S}^F(D)$  the $F-$fold sym\-me\-tric product of $D$,
there are multiline\-ar,  $B-$equiva\-riant  ({\it i.e.} which respect 
the action  of  $B$) maps
 $\left\{~~, \cdots,~~ \right\}$ from 
${\cal S}^F\left(A_k\right)$ into  $B$.
In other words, we  assume that some of the elements of the Lie algebra $B$ can
be expressed as $F-$th order symmetric products of
``more fundamental generators''. 
\end{enumerate}

\noindent
Of course, to ensure the coherence of this algebraic structure more 
is needed (Jacobi identities, unitarity,  representations, {\it etc.})
as can be seen in \cite{vg}.
From the above definition of  $F-$Lie algebras, firstly,
we observe that no relation between different graded sectors is 
postulated. Secondly, the
sub-space  $B \oplus A_1 \subset S$ satisfies all the conditions of $F-$Lie
algebras. 

\section{Fractional supersymmetry}
Along the lines of $F-$Lie algebras it has then been 
shown that it is possible  to
define fractional supersymmetry (FSUSY) starting from any Lie algebra $\g$ 
and any representation $\D$. The basic idea is the following. Let $\g$ be a
rank $r$ semi-simple Lie algebra and denote $\alpha_i$ the set of 
primitive roots and
$\mu_i$ the set of fundamental weight vectors ($2 \frac{\mu_i . \alpha_j}
{(\alpha_j)^2} = \delta_{ij}$). If one considers 
$B=\g \oplus \D_\mu$
($\D_\mu$ being the representation associated to the 
primitive vector $|\mu>$ with weight $\mu= \sum \limits_{i=1}^r n_i \mu_i,
n_i \in \NN$ --highest weight 
representation--)  as
the bosonic part of the $F-$Lie algebra (as a semi-direct product)

\begin{equation}
\label{eq:bos}
\left[ \g, \g \right] \subset \g,\ \  
\left[ \g, \D_\mu \right] \subset \D_\mu,
\ \  \left[ \D_\mu, \D_\mu \right]=0,
\end{equation}

\noi
one possible solution for the
grade one sector is to  consider the infinite dimensional representation
$\D_{\mu/F},$ with highest weight of weight $\mu/F=\sum \limits_{i=1}^r 
\frac{n_i}{F}  \mu_i,
n_i \in \NN$. The action of
$\g$ on $\D_{\mu/F}$ is extended as follow

\begin{equation}
\label{eq:bos_rrep}
 \left[ \g, \D_{\mu/F} \right] \subset \D_{\mu/F},
\ \  \left[ \D_\mu, \D_{\mu/F} \right]=0.
\end{equation}

\noi
 The representation $\D_{\mu/F}$
 is in general infinite dimensional and non-unitary and is called a Verma
module \cite{kr}. 
In an earlier  paper an FSUSY associated to $\g, \D_\mu$ and $\D_{\mu/F}$
was constructed in an abstract way \cite{vg}. It was then noticed that  
for any Lie algebras it is possible to construct a differential realization
and to obtain any representation in terms of appropriate homogeneous 
monomials \cite{fliealg}. In such a realization FSUSY can be explicitly 
obtained by action on appropriate homogeneous monomials. Roughly speaking,
in FSUSY the representations $\D_\mu$ and $\D_{F/F}$ can be related

\begin{equation}
\label{eq:fsusy}
{\cal S}^F \left(\D_{\mu/F} \right) \sim \D_\mu,
\end{equation} 

\noi
but the main difficulty, in such a 
construction, is connected to the requirement of 
relating an infinite dimensional
representation $\D_{\mu/F}$ to a finite dimensional one $\D_\mu$ 
in an equivariant
way, {\it i.e.} respecting the action of $\g$. One possible way of 
solving this contradiction is to embed $\D_\mu$ into an infinite dimensional
(reducible but indecomposable) representation \cite{vg} \cite{fliealg}.
Another possibility is to embed $\g$ into a infinite dimensional algebra 
dubbed $V(\g)$ \cite{vg}.
   
\section{Fractional supersymmetry and $\s(3,\RR)$}

Here, our purpose is to analyze these two points in more details and in 
particular
to construct explicitly the infinite dimensional algebra $V(\g) \supset \g$.
The first point was already analyzed in \cite{vg} \cite{fliealg} and 
the differential
realizations was the main issue in \cite{fliealg}.
Indeed, one additional  advantage of the differential realizations  given in 
\cite{fliealg} is   to provide the main tool to obtain $V(\g)$ from $\g$. 
Up to now, the solution has been obtained only for $\g=\s(n, \RR)$ (the
other series are under investigation).
Furthermore, the formulas are generic for all $\s(n,\RR)$ so we may
just focus here  on $\g=\s(3,\RR)$. 

\subsection{Fractional supersymmetry and finite dimensional algebras}
\medskip

Let $\alpha, \beta$ be the simple roots of $\s(3,\RR)$,
$\gamma=\alpha+\beta$ the third positive root and  
$\mu_1$ and $\mu_2$ the fundamental weights.
Introduce,  $x_1,x_2,x_3 >0$  
of weight $|\mu_{1}> = x_1, |\mu_{1}-\alpha>=x_2, 
|\mu_{1}-\alpha -\beta>=x_3$. This means that $\D_{\mu_1}=<x_1,x_2,x_3>$ is
just the three dimensional fundamental representation.
From the definition of this representation and from its explicit construction
it follows that   the $\s(3,\RR)$ generators take the form (the 
normalization taken here is not conventional, but is most
 useful for the sequel)

\begin{eqnarray}
\label{eq:su3}
&& \hskip-.8truecm
\begin{array}{ll}
E^\alpha = x_1 \partial_{x_2},&E^{-\alpha} = x_2 \partial_{x_1}, \cr
E^\beta = x_2 \partial_{x_3},&E^{-\beta} = x_3 \partial_{x_2}, \cr
E^\gamma = x_1 \partial_{x_3},&E^{-\gamma} = x_3 \partial_{x_1}, 
\end{array} \\
&& \hskip-.8truecm
\ h_1=x_1 \partial_{x_1} - x_2 \partial_{x_2}, \nonumber \\
&& \hskip-.8truecm
 \ h_2=x_2 \partial_{x_2}  - x_3 \partial_{x_3}. \nonumber
\end{eqnarray}

\noindent
Following the results established in \cite{fliealg} all representations
of $\s(3,\RR)$ can then be obtained from the fundamental representation
and consequently we only need one differential realization to obtain
all the representations of  $\s(3,\RR)$.  Indeed,
the highest weight of
the representation $\D_{ p_1 \mu_1 + p_2 \mu_2}$ is simply
$|p_1 \mu_1 + p_2 \mu_2>=
\big(x_1\big)^{p_1} \big(x_1 \wedge x_2\big)^{p_2}$ because
(i) $E^{\alpha,\beta} |p_1 \mu_1 + p_2 \mu_2> =0$ and (ii)
$h_i |p_1 \mu_1 + p_2 \mu_2>= p_i |p_1 \mu_1 + p_2 \mu_2>$.
The representation $\D_{p_1 \mu_1 + p_2 \mu_2}$ is then  obtained from the
action on $|p_1 \mu_1 + p_2 \mu_2>=
\big(x_1\big)^{p_1} \big(x_1 \wedge x_2\big)^{p_2}$ of $E^{-\alpha},
E^{ -\beta}$
given in (\ref{eq:su3}). A direct calculation shows that when
$p_1, p_2$ are integer the representation is finite dimensional,
and the  operators $E^{\pm \alpha}, E^{\pm \beta}, E^{\pm \gamma}$ are
nilpotent. However, when at least one of the $p_i$ is not an 
integer the representation
is infinite dimensional. As  shown in \cite{fliealg} this 
procedure can be equally applied for any Lie algebras.

The aim is now to construct an $F-$Lie algebra associated to the three
dimensional representation $\D_{\mu_1}$
(this could have been done for any representation of  $\s(3,\RR)$). 
In the  realization (\ref{eq:su3})
the vectorial representation writes 

\begin{eqnarray}
\D_{\mu_{1}}= \left\{
\begin{array}{lll}
x_1&\hskip -.2truecm=|\mu_{1}>,& \cr
x_2&\hskip -.2truecm= |\mu_{1}-\alpha>
&\hskip -.2truecm=E^{-\alpha}|\mu_{1}>, \cr 
x_3&\hskip -.2truecm=|\mu_{1}-\alpha -\beta>&
\hskip -.2truecm=E^{-\beta}E^{-\alpha}|\mu_{1}>. 
\end{array}  \nonumber
\hskip-.7truecm
\right. \nonumber
\end{eqnarray}  
\begin{equation}
\label{eq:3}
\end{equation}

\noindent
Hence, 
\begin{equation}
\label{eq:bos2}
B=\s(3,\RR) \oplus \D_{\mu_{1}}
\end{equation}

\noi
is considered
for the bosonic (graded zero part) of the $F-$Lie algebra. The natural
representation to define the ``$\mathrm{F}^{\mathrm{th}}-$root'' of 
$\D_{\mu_{1}}$ is  $\D_{\mu_{1}/F}$.
So,  the graded one part is taken as

\begin{equation}
\label{eq:anyon}
A_1 = \D_{\frac{\mu_{1}}{F}}.
\end{equation}

\noindent
The primitive vector of this representation is given by $x_1^{1/F}$.
In the differential realization (\ref{eq:su3}), this representation, acting 
explicitly with the operators $E^{-\alpha},E^{-\beta}$ 
on $x_1^{1/F}= |\frac{\mu}{F}>$  is easily obtained

\begin{eqnarray}
\label{eq:3/F}
&& \hskip -.7truecm
\D_{\frac{\mu_{1}}{F}} = \Big \{
|\frac{\mu_{1}}{F} - n \alpha - p \beta> = \nonumber \\
&&  \hskip -.7truecm
(x_1)^{1/F} \left(\frac{x_2}{x_1}\right)^n  
 \left(\frac{x_3}{x_2}\right)^p  \\
&& \hskip -.7truecm
\sim \left(E^{-\beta}\right)^p \left(E^{-\alpha}\right)^n 
|\frac{\mu_{1}}{F}>, 
n \in \NN, 0 \le p \le n  \Big\}, \nonumber
\end{eqnarray} 
\noi
leading to the weight diagram given in figure 1.\\

\begin{figure*}
\unitlength1cm
\begin{picture}(5,7.3)
\put(11.7,6.8){$x_1^{1/F}$} 
\put(9.8,5.1){$x_1^{1/F}\left(\frac{x_2}{x_1}\right)$}
\put(8,3.3){$x_1^{1/F}\left(\frac{x_2}{x_1}\right)^2$}
\put(6.2,1.4){$x_1^{1/F}\left(\frac{x_2}{x_1}\right)^3$}
\put(8.4,6.8){$x_1^{1/F}\left(\frac{x_3}{x_1}\right)$}
\put(6.5,5.1){$x_1^{1/F}\left(\frac{x_3}{x_1}\right)
\left(\frac{x_2}{x_1}\right)$}
\put(4.5,3.3){$x_1^{1/F}\left(\frac{x_3}{x_1}\right)
\left(\frac{x_2}{x_1}\right)^2$}
\put(4.8,6.8){$x_1^{1/F}\left(\frac{x_3}{x_1}\right)^2$}
\put(2.8,5.1){$x_1^{1/F}\left(\frac{x_3}{x_1}\right)^2
\left(\frac{x_2}{x_1}\right)$}
\put(1.,6.8){$x_1^{1/F}\left(\frac{x_3}{x_1}\right)^3$}
\put(0,0){\epsffile{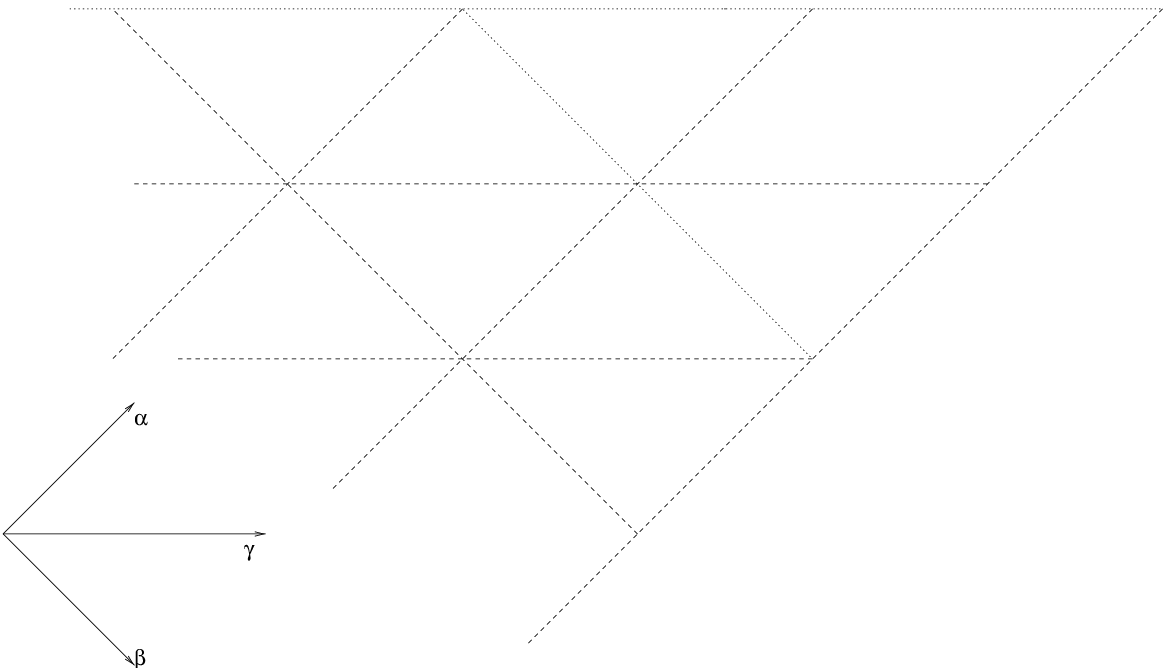}} 
\end{picture}
\caption{\small{{Weight diagram for the $\D_{\mu_{1}/F}$ representation of
$\s(3,\RR)$. The down-left corner represents the positive roots of $\s(3,\RR)$.
The representation is  infinite dimensional in the direction of $\alpha$
and $\gamma$, but finite dimensional in the direction $\beta$.
Two consecutive vectors in the direction $\alpha$ (resp. $\beta$ or
$\gamma$) are related by a multiplication by $\frac{x_2}{x_1}$
(resp.$\frac{x_3}{x_2}$ or $\frac{x_3}{x_1}$).}}}
\end{figure*}

Next to define an $F-$Lie algebra associated to 
$\s(3,\RR)\oplus\D_{\mu_{1}} \oplus \D_{\mu_{1}/F}$   the following
representation (reducible) is considered

\begin{eqnarray}
\label{eq:sf}
&&\hskip-.8truecm
{\cal S}^F\left(\D_{\mu_{1}/F}\right)= 
\left\{ \odot_{i=1}^F (x_1)^{1/F} \left(\frac{x_2}{x_1}\right)^{n_i}
\nonumber \right.\\
&&\hskip-.8truecm
\left. \left(\frac{x_3}{x_2}\right)^{p_i}, \ \ n_i \in \NN, 0 \le p_i \le n_i
 \right\},
\end{eqnarray}

\noi
with $\odot$ the symmetric tensorial product.

Now, comparing $x_1$, the primitive vector of $\D_{\mu_{1}}$, with 
$\otimes^F\left(x_1\right)^{1/F} \in {\cal S}^F\left(\D_{\mu_{1}/F}\right)$ 
it is seen that these two vectors,
as primitive vectors, satisfy the same properties

\begin{eqnarray}
\begin{array}{ll}
h_1(x_1)=x_1,&h_1\Big(\otimes^F\left(x_1\right)^{\frac{1}{F}}\Big)=
\otimes^F\left(x_1\right)^{\frac{1}{F}}, \cr
h_2(x_1)=0,&h_2\Big(\otimes^F\left(x_1\right)^{\frac{1}{F}}\Big)=0, \cr
E^{\alpha,\beta,\gamma}(x_1)=0,&E^{\alpha,\beta,\gamma}
\Big(\otimes^F\left(x_1\right)^{\frac{1}{F}}\Big)=0.
 \end{array}  \nonumber
\end{eqnarray}
\begin{equation}
\label{eq:primitive}
\end{equation}

But now, constructing the representation from these primitive vectors,
one obtains, on the one hand  $\D_{\mu_{1}}$, and on the other hand,
the infinite dimensional representation

\begin{eqnarray}
\label{eq:verma}
&&\hskip -.6truecm
\left<\otimes^F\left(x_1\right)^{1/F} \right> = \Big\{ 
 |\mu_{1} - n \alpha - p \beta> =  \hskip-1.5truecm\nonumber  \\ 
&& \hskip -.6truecm
\left(E^{-\beta}\right)^p \left(E^{-\alpha}\right)^n 
\Big(\otimes^F\left(x_1\right)^{1/F}\Big), 
n \in \NN, 0 \le p \le n  
\Big \}. \nonumber \\ \hskip -1.truecm
\end{eqnarray}

\noi
However, a direct calculation shows that the following relations hold
\begin{eqnarray}
\label{eq:nil}
&&\hskip -.7truecm 
E^\alpha |\mu_{1}- 2 \alpha> =0 \nonumber \\
&&\hskip -.7truecm 
E^\alpha |\mu_{1}- 2 \alpha - \beta> =0 \hfill\hskip .7truecm    \\
&&\hskip -.7truecm 
E^\gamma |\mu_{1}- 2 \alpha - 2 \beta> =0.  \nonumber
\end{eqnarray}

\noi
It means that ${\cal V}_{\mu_{1}}=\left<\otimes^F\left(x_1\right)^{1/F} 
\right>$ is a Verma module
(the operator $E^\alpha$ is not nilpotent). The relation with the finite 
dimensional  representation $\D_\mu$ is the following isomorphism
\begin{equation}
\label{eq:verma_rep}
D_\mu \cong \left<\otimes^F\left(x_1\right)^{1/F} \right>/{\cal M}_\mu
\end{equation}

\noi
with ${\cal M}_\mu=\Big\{\left(E^{-\beta}\right)^p \left(E^{-\alpha}\right)^n
\big(\otimes^F\left(x_1\right)^{1/F}\big),$
$n \in \NN, 0 \le p \le n, (n,p) \ne (0,0), (1,0), (1,1) \Big\}$  the
maximal proper sub-representation of ${\cal V}_{\mu_{1}}$
($\forall t \in \s(3, \RR), \forall m \in {\cal M}_\mu, t(m) \in{\cal  M}_\mu$)
(see \cite{fliealg} for more details).\\

%
%


Thus, 
to obtain an $F-$Lie algebra some constraints have to be introduced.
Following \cite{vg}, ${\cal F}$ is defined as the vector space of functions
on  $x_1,x_2,x_3 >0$. The multiplication map $m_n: {\cal F} \times 
\cdots \times {\cal F} \longrightarrow {\cal F}$ given by
$m_n(f_1,\cdots,f_n) = f_1 \cdots f_n$ is multilinear and totally symmetric.
Hence, it induces a map $\mu_F$ from ${\cal S}^F\left({\cal F} \right)$
into ${\cal F}$. Restricting to ${\cal S}^F
\left({\D_{\frac{\mu_{1}}{F}}} \right)$ one
gets $\mu_F:{\cal S}^F\left({\D_{\frac{\mu_{1}}{F}}} \right) \longrightarrow
{\cal S}_{{\mathrm {red}}}^F\left({\D_{\frac{\mu_{1}}{F}}} \right)$


\begin{eqnarray}
\label{eq:sfred}
&&\hskip-.65truecm
\mu_F \left(\odot_{i=1}^F (x_1)^{1/F} \left(\frac{x_2}{x_1}\right)^{n_i}
\left(\frac{x_3}{x_2}\right)^{p_i}\right) \nonumber \\
&&\hskip-.65truecm
= 
x_1 \left(\frac{x_2}{x_1}\right)^{\sum \limits_{i=1}^F n_i}
\left(\frac{x_3}{x_2}\right)^{\sum \limits_{i=1}^F p_i}.
\end{eqnarray}

\noi
It is easily seen that 
${\cal S}_{{\mathrm {red}}}^F\left({\D_{\frac{\mu_{1}}{F}}} \right)
=\left\{
x_1 \left(\frac{x_2}{x_1}\right)^n  \left(\frac{x_3}{x_2}\right)^p, 
n \in \NN, 0 \le p \le n  \right\}
 \supset
\D_{\mu_{1}}$.
This representation is reducible but indecomposable.
Namely, a complement of $\D_{\mu_{1}}$ in 
${\cal S}_{{\mathrm {red}}}^F\left({\D_{\frac{\mu_{1}}{F}}} \right)$, stable 
under $\s(3,\RR)$ cannot be found: {\it i. e.} we cannot write
${\cal S}_{{\mathrm {red}}}^F\left({\D_{\frac{\mu_{1}}{F}}} \right) =
\D_\mu \oplus V$ in an equivariant way (respecting the action of  $\s(3,\RR)$).
For instance $x_1 \left(\frac{x_2}{x_1}\right)^2 \in V$ is such
that $E^\alpha \left(x_1 \left(\frac{x_2}{x_1}\right)^2\right)= 
2 x_2 \in \D_\mu$,
but $E^{-\alpha}(x_2)=0$.
%
%
With this simple observation the conclusion is, as in \cite{vg}, that
 an $F-$Lie algebra with $B=\s(3,\RR) \oplus \D_{\mu_{1}}$ cannot be defined.
To obtain such a structure, one possible solution is
to embed $\D_{\mu_{1}}$ into an infinite dimensional representation. 
For instance,
 
\begin{eqnarray}
\label{eq:flie}
&&\hskip-.7truecm\left(\s(3,\RR) \oplus 
{\cal S}^F\left( {\cal D}_{\frac{\mu_{1}}{F}}\right)_{\mathrm {red}}\right)
\oplus {\cal D}_{\frac{\mu_{1}}{F}} \nonumber \\
&&\hskip-.7truecm
\supset \left(\s(3, \RR) \oplus \D_\mu \right) \oplus {\cal D}_{\frac{\mu_{1}}{F}}\
\end{eqnarray}

\noi
has a structure of $F-$Lie algebras. See \cite{vg} and \cite{fliealg} for more 
details.

\subsection{Fractional supersymmetry and infinite dimensional algebras}
As  mentioned there is another way to define an $F-$Lie 
algebra, extending $\g$ into an infinite dimensional Lie algebra. 
To define this algebraic structure,
the key observations are as follow:

(i) For any Lie algebra $\g$ (complexified)
and to any positive roots of $\g$ there is a natural $\s(2,\RR)$ sub-algebra
generated by $\{E^{\pm \alpha}, h_\alpha\}$ and fulfilling the commutation 
relations $[h_\alpha, E^{\pm \alpha}]=\pm E^{\pm \alpha}, \ \ [E^\alpha, 
E^{-\alpha}] = 2 h_\alpha$.

(ii)  The centerless Virasoro algebra (generated by  $L_n, n\in \NN$
with $[L_n, L_m] = (n-m) L_{n+m}$) contains an $\s(2,\RR)$ sub-algebra,
generated by $\{L_{\pm 1}, L_0 \}$.

(iii)  Moreover, for all $n >0$ the generators $\{L_{\pm n}, L_0\}$
 of the Virasoro algebra generate a $\s(2,\RR)$ sub-algebra.

Putting (i) and (ii-iii) together, the definition of $V(\g)$
is given, as a possible generalization of the centerless Virasoro algebra
(de Witt algebra). This construction, proceeds in two steps.
Firstly a partial Lie  algebra is defined.
Secondly, calculating all the commutators of all the generators defined
in the first step, the algebra is closed leading to $V(\g)$.

\subsubsection{The partial Lie algebra}
\medskip

Having (i-iii) in mind, the following is assumed, in order to obtain
a possible extension of the Virasoro (centerless) algebra 
\begin{itemize}
\item For any positive roots  $\alpha$ there is a Virasoro algebra,
namely generators of weight $n \alpha, n \in \ZZ$.
\item For all $n >0$ they are generators of weights $\pm n \alpha, 
\pm n \beta, \pm n \gamma$ closing, with $h_1,h_2$ through an $\s(3,\RR)$
algebra.
\end{itemize}

\noi
Such an algebra was defined in a formal and universal way in \cite{vg}.
But now, the relevance of   the variables $x_1,x_2,x_2$
is clear, in the sense
that $\left( \frac{x_1}{x_2}\right)^n$ is of weight $n \alpha$ and 
$\left( \frac{x_2}{x_3}\right)^n$ of weight $n \beta$
and contain all the informations to construct generators of appropriate
weight. It implies that
the only generators of weight $n \alpha$ are 
$\left( \frac{x_1}{x_2}\right)^n x_i \partial_{x_i}$, 
of weight $n \beta$
are $\left( \frac{x_2}{x_3}\right)^n x_i \partial_{x_i}$, and of
weight $n \gamma$ are 
$\left( \frac{x_1}{x_3}\right)^n x_i \partial_{x_i}$, with 
$i=1,2,3$. (Strictly speaking $(x_1 x_2 x_3)$ being of weight zero,
more  generators can be defined.)
Finally,  introducing (for $n>0$)

\begin{eqnarray}
\begin{array}{ll}
E^{n \alpha}= \left(\frac{x_1}{x_2}\right)^n x_2 \partial_{x_2},&
E^{-n \alpha}= \left(\frac{x_2}{x_1}\right)^n x_1 \partial_{x_1} \cr
E^{n \beta}= \left(\frac{x_2}{x_3}\right)^n x_3 \partial_{x_3}, &
E^{-n \beta}= \left(\frac{x_3}{x_2}\right)^n x_2 \partial_{x_2},\cr
E^{n \gamma}= \left(\frac{x_1}{x_3}\right)^n x_3 \partial_{x_3},&
E^{-n \gamma}= \left(\frac{x_3}{x_1}\right)^n x_1\partial_{x_1}, 
\end{array}
\nonumber
\end{eqnarray}
\begin{equation}
\label{eq:partial}
\end{equation}
\begin{eqnarray}
\begin{array}{ll}
\tilde E^{n \alpha}= \left(\frac{x_1}{x_2}\right)^n x_1\partial_{x_1}, &
\tilde E^{-n \alpha}= \left(\frac{x_2}{x_1}\right)^n x_2 \partial_{x_2}, \cr
\tilde E^{n \beta}= \left(\frac{x_2}{x_3}\right)^n x_2\partial_{x_2}, &
\tilde E^{-n \beta}= \left(\frac{x_3}{x_2}\right)^n x_3 \partial_{x_3},\cr
\tilde E^{n \gamma}= \left(\frac{x_1}{x_3}\right)^n x_1 \partial_{x_1},&
\tilde E^{-n \gamma}= \left(\frac{x_3}{x_1}\right)^n x_3\partial_{x_3},
\end{array}
\nonumber
\end{eqnarray}

\noi
a direct and simple calculation shows that these generators 
(with $h_1,h_2$ (see (\ref{eq:su3})) and $T= x_1 \partial_{x_1} + x_2 \partial_{x_2} + 
x_3 \partial_{x_3}$), endow the following structure:

\begin{enumerate}
\item For all $n >0$, $\left\{ E^{\pm n  \alpha},  E^{\pm n \beta},
 E^{\pm n \gamma},
h_1,h_2 \right\}$ and  $\left\{\tilde E^{\pm n \alpha},\tilde E^{\pm n \beta},
\tilde E^{\pm n \gamma},h_1,h_2 \right\}$ generate an $\s(3, \RR)$ algebra.
\item For all positive root $\varphi=\alpha,\beta,\gamma$,
$\left\{  E^{-n \varphi}, \tilde E^{n \varphi}, n \in \NN, 
\frac{1}{2}[\tilde E^\varphi, E^{-\varphi}]
\right \}$ and 
$\left\{\tilde E^{-n \varphi},  E^{n \varphi},\right.$
$\left.  n \in \NN,
\frac{1}{2}[E^\varphi,\tilde E^{-\varphi}] \right \}$ generate a 
Virasoro 
algebra.
\end{enumerate}

\noi
It means that the partial Lie algebra so-constructed admits a concentric 
 and a radial symmetry. Concentric symmetries are related to
$\s(3,\RR)$'s sub-algebras and radial ones to Virasoro's sub-algebras.
All this is summarized in Figure 2.

 \begin{figure*}
\unitlength1cm
\begin{picture}(5,8.)
\put(5.5,7.5){$E^{n \alpha}= \left(\frac{x_1}{x_2}\right)^n x_2 \partial_{x_2} $}
\put(5.5,6.5){$\tilde E^{n \alpha}= \left(\frac{x_1}{x_2}\right)^n x_1 
\partial_{x_1} $}
\put(1.5,2.){$E^{-n \alpha}= \left(\frac{x_2}{x_1}\right)^n x_1 
\partial_{x_1} $}
\put(1.5,1){$\tilde E^{-n \alpha}= \left(\frac{x_2}{x_1}\right)^n x_2 
\partial_{x_2} $}
\put(5.5,2.){$E^{n \beta}= \left(\frac{x_2}{x_3}\right)^n x_3 
\partial_{x_3} $}
\put(5.5,1.) {$\tilde E^{n \beta}= \left(\frac{x_2}{x_3}\right)^n x_2
\partial_{x_2} $}
\put(1.5,7.5){$E^{-n \beta}= \left(\frac{x_3}{x_2}\right)^n x_2 
\partial_{x_2} $}
\put(1.5,6.5) {$\tilde E^{-n \beta}= \left(\frac{x_3}{x_2}\right)^n x_3 
\partial_{x_3} $}
\put(8,4.45){$E^{n \gamma}= \left(\frac{x_1}{x_3}\right)^n x_3 
\partial_{x_3} $}
\put(8,3.45){$\tilde E^{n \gamma}= \left(\frac{x_1}{x_3}\right)^n x_1 
\partial_{x_1} $}
\put(0.,4.45){$E^{-n \gamma}= \left(\frac{x_3}{x_1}\right)^n x_1 
\partial_{x_1} $}
\put(0.,3.45){$\tilde E^{-n \gamma}= \left(\frac{x_3}{x_1}\right)^n x_3
\partial_{x_3} $}
%
\put(4.,4.7){$h_\alpha= x_1  \partial_{x_1} - x_2  \partial_{x_2}$}
\put(3.5,4.2){$T= x_1  \partial_{x_1} + x_2  \partial_{x_2} + x_3  
\partial_{x_3}$} 
\put(4.,3.7){$h_\beta=  x_2  \partial_{x_2} - x_3  \partial_{x_3}$}
%
\put(1.,0){direction $ \ \ \alpha$}
\put(6.5,0){direction $ \ \ \beta$}
\put(12.,4.2){direction $ \ \ \gamma$}
\put(0,0){\epsffile{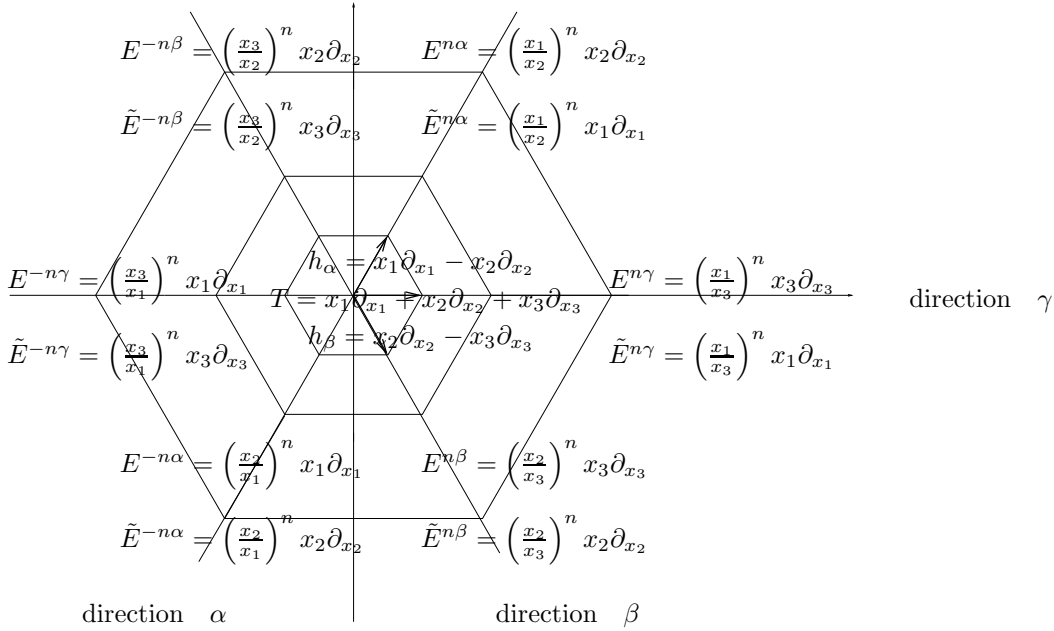}} 
\end{picture}
\caption{
The partial Lie algebra of $V[\s(3,\RR)].$
For all $n > 0$ there are two $\s(3,\RR)$ algebras (concentric  symmetry). 
For all positive roots  two centerless Virasoro algebras (radial
symmetry) occur.}
\end{figure*}

\subsubsection{The algebra $V[\s(3,\RR)]$}
\medskip

With the above introduced generators of the partial Lie algebra given in
(\ref{eq:partial}) and upon calculating the commutators that are neither on the
same circle, nor on the same radial,  the generators of
$V[\s(3,\RR)]$ are obtained

\begin{eqnarray}
\label{eq:Vg}
X^{n,m}_1&=& \left(\frac{x_1}{x_2}\right)^n \left(\frac{x_2}{x_3}\right)^m
x_1 \partial_{x_1}, \nonumber \\
X^{n,m}_2&=& \left(\frac{x_1}{x_2}\right)^n \left(\frac{x_2}{x_3}\right)^m
x_2 \partial_{x_2}, \\
X^{n,m}_3&=& \left(\frac{x_1}{x_2}\right)^n \left(\frac{x_2}{x_3}\right)^m
x_3 \partial_{x_3}. \nonumber
\end{eqnarray}

\noi
Denoting the generators in a generic way $X^{n,m}_i = 
(x_1)^{p_1}(x_2)^{p_2}(x_3)^{p_3} x_i \partial_{x_i}$ with $p_1 =n, p_2 =m-n,
p_3=-m$ ($p_1+p_2+p_3=0$), the commutation relations take the form


\begin{eqnarray}
 [X^{n,m}_i, X^{n^\prime,m^\prime}_j]= 
 p^\prime_i X^{n+n^\prime, m+ m^\prime}_j
-p_j X^{n+n^\prime, m+ m^\prime}_i. \nonumber
\end{eqnarray}
\begin{equation}
\label{eq:CR}
\end{equation}

\noi
It can  be noticed that this algebra admits a Lie algebra  automorphism.
Indeed, a direct calculation shows that  if $X^{n,m}_i \longrightarrow 
\rho(X^{n,m}_i)$ such that  $\rho(X^{n,m}_1+ X^{n,m}_2+X^{n,m}_3)=0$,
the commutation relations (\ref{eq:CR}) remains  unchanged. The interpretation
of the obtained  algebra, as well as of the automorphism $\rho$ is left
for a future publication. Notice only that the algebra
$V[\s(2,\RR)]/{\mathrm Ker} \rho$ is just the Virasoro algebra and
 that the weight diagram of $V[\s(3,\RR)]$ covers
all points $n \alpha + m \beta, n, m \in \ZZ$ of the 
weight lattice  with a degeneracy of three.\\

From the definition of the Lie algebra $V[\s(3,\RR)]$ it is not difficult 
to extend the 
representations (\ref{eq:3}) and (\ref{eq:3/F}). Using (\ref{eq:Vg}) and
starting with (\ref{eq:3}) and (\ref{eq:3/F}), one gets
\begin{eqnarray}
\label{eq:rep_Vg}
&& \hskip -.7truecm
\hat \D_{\mu_{1}}= \Big \{|\mu_1 + r \alpha + s \beta > =
\nonumber \\ 
&& \hskip -.7truecm  x_1 
\left(\frac{x_1}{x_2}\right)^r \left( \frac{x_2}{x_3} \right)^s,  
 r,s \in \ZZ \Big\}  
\nonumber \\
\\
&& \hskip -.7truecm
\hat \D_{\frac{\mu_{1}}{F}}=\Big \{ 
|\frac{\mu_1}{F} + r \alpha + s \beta > = 
\nonumber \\
&& \hskip -.7truecm
(x_1)^{1/F}
\left(\frac{x_1}{x_2}\right)^r \left( \frac{x_2}{x_3} \right)^s, 
r,s \in \ZZ \Big\} \nonumber
\end{eqnarray}

\noi 
and 

\begin{eqnarray}
\label{eq:rep}
&&\hskip -.7truecm
X_1^{n,m} |\mu_1 + r \alpha +s \beta> \nonumber \\
&&\hskip -.7truecm
=(1+r) |\mu_1 +(r+n) \alpha + (s+m) \beta > \nonumber \\
&&\hskip -.7truecm
X_2^{n,m} |\mu_1 + r \alpha +s \beta>   \nonumber \\ 
&&\hskip -.7truecm
=(s-r) |\mu_1 +(r+n) \alpha + (s+m) \beta > \nonumber \\ 
&&\hskip -.7truecm
X_3^{n,m} |\mu_1 + r \alpha +s \beta>\nonumber \\
&&\hskip -.7truecm
 = -s |\mu_1 +(r+n) \alpha + (s+m) \beta>  \nonumber \\
\\
&&\hskip -.7truecm
X_1^{n,m} | \frac{\mu_1}{F} + r \alpha + s \beta>  \nonumber \\
&&\hskip -.7truecm
= (\frac{1}{F}+r) | \frac{\mu_1}{F} + (n+r) \alpha + (m+s) \beta > 
\nonumber \\
&&\hskip -.7truecm
X_2^{n,m} | \frac{\mu_1}{F} + r \alpha + s \beta>  \nonumber \\
&&\hskip -.7truecm
=  (s-r) | \frac{\mu_1}{F} + (n+r) \alpha + (m+s) \beta > \nonumber \\
&&\hskip -.7truecm
X_3^{n,m} | \frac{\mu_1}{F} + r \alpha + s \beta>  \nonumber \\
&&\hskip -.7truecm
= -s  | \frac{\mu_1}{F} + (n+r) \alpha + (m+s) \beta   > \nonumber
\end{eqnarray}

\noi
It may be seen that the representations given in (\ref{eq:3}) and 
(\ref{eq:3/F}) are included in their corresponding representations 
given in (\ref{eq:rep_Vg}). 
It should also be noted that these two representations are not obtained from
a primitive vector {\it i.e.} they are not bounded from bellow or above.

Furthermore,  in each case the action of 
$V[\s(3,\RR)]$ 
extends the action of $\s(3,\RR)$. 
The fundamental property of these representations lies in
a $V[\s(3,\RR)]-$equivariant map from 
${\cal S}^F\left({\hat \D_{\frac{\mu_{1}}{F}}}\right)$ into $\hat \D_{\mu_1}$.
This is just the multiplication map $\mu_F$ (see (\ref{eq:sfred})).
 A direct calculation shows that 

\begin{eqnarray}
\label{eq:red_vg}
&&\hskip-.6truecm
{\cal S}_{{\mathrm {red}}}^F\left({\hat \D_{\frac{\mu_{1}}{F}}}\right)
 \buildrel{\hbox{def}} \over =
\mu_F\left(
{\cal S}^F\left({\hat \D_{\frac{\mu_{1}}{F}}}\right)
\right) \\
&&\hskip-.6truecm
=\left\{x_1 \left(\frac{x_1}{x_2}\right)^r \left(\frac{x_2}{x_3}\right)^s,
\ \ r,s \in \ZZ \right\} \cong \hat \D_{\mu_1} \nonumber
\end{eqnarray} 

\noi
and then 
${\cal S}_{{\mathrm {red}}}^F\left({\hat \D_{\frac{\mu_{1}}{F}}}\right)$
and $\hat \D_{\mu_1}$ are isomorphic. Hence, 
\begin{equation}
\label{eq:flie2}
\Big(V[\s(3,\RR)] \oplus  \hat \D_{\mu_1} \Big) \oplus\hat  \D_{\frac{\mu_1}{F}}
\end{equation}

\noi 
is an $F-$Lie algebra.\\

\section{Conclusion}
To conclude, 
considering infinite dimensional representations,
one is able to construct a theory generalizing supersymmetry and
super-Lie algebras. Indeed, starting from a Lie algebra $\g$ and a 
representation
$\D_\mu$, the basic point is  to consider  the infinite dimensional
representation $\D_{\mu/F}$. As shown this  construction leads
naturally to an infinite dimensional algebra $V(\g)$ containing $\g$ as a
sub-algebra. When $\g=\s(2,\RR)$ this algebra reduces to the centerless 
Virasoro algebra. This construction was done explicitly
for $\g=\s(3,\RR)$, but such a construction works equally well for 
$\g= \s(n,\RR)$.
It should then be interesting to study this algebra {\it per se}:
geometrical interpretation, central extension {\it etc.}.
Indeed, it was then realized that the algebra $V(\s(3,\RR))/Ker \rho$ is
included into the algebra of vector fields on the Torus $T^2$.
Let us mention that a general study is under investigation. 
Following the same simple principle, we are looking  for the possibility to
realize  some Lie algebras (including  Kac-Moody and 
hyperbolic algebras \cite{k}) as a sub-algebra of the vector fields on a
torus \cite{rs}.
Finally, let us mention that for $\g=\s(2,\RR)$ unitary 
representations of $(1+2)D$ FSUSY have been constructed \cite{fsusy3d}. As
observed earlier in this case FSUSY is a 
symmetry which acts on relativistic anyons
\cite{an}. The interpretation of FSUSY in higher dimensional 
space-time is still an open question.

\end{document}